\documentclass{aa}

\usepackage{graphicx}
\usepackage{txfonts}

\begin{document}

	\title{Herbig-Haro flows in L1641N
		\thanks{Based on observations made with the Nordic Optical Telescope, operated on the island of
			La Palma jointly by Denmark, Finland, Iceland, Norway, and Sweden, in the Spanish Observatorio
			del Roque de los Muchachos of the Instituto de Astrofisica de Canarias.}\fnmsep
		\thanks{This work is based in part on observations made with the Spitzer Space Telescope, which is operated
			by the Jet Propulsion Laboratory, California Institute of Technology under a contract with NASA.}
	}

	\author{M.~G\aa lfalk \inst{1} \and
		G.~	Olofsson \inst{1}
		}
    
	\offprints{\\M. G\aa lfalk, \email{magnusg@astro.su.se}}

	\institute{Stockholm Observatory, Sweden				
	      }
	\date{Received 01 December 2006 / Accepted 06 February 2007}

\abstract
{}
{To study the Herbig-Haro (HH) flows in L1641N, an active star formation region in the southern part of the Orion GMC 
and one of the most densely populated regions of HH objects in the entire sky. By mapping the velocities of these HH
objects, combined with mid-IR observations of the young stars, the major flows in the region and the corresponding outflow sources
can be revealed.\vspace{2mm}
}			
{We have used the 2.56\,m Nordic Optical Telescope (NOT) to observe two deep fields in L1641N, selected on the basis of previous
shock studies, using the  2.12\,$\mu$m transition of H$_2$ (and a $K_S$ filter to sample the continuum) for a total
exposure time of 4.6\,hours (72\,min $K_S$) in the overlapping region. The resulting high-resolution mosaic
(0.23\,$\arcsec$ pixel size, 0.75\,$\arcsec$ seeing) shows numerous new shocks and resolves many known
shocks into multiple components. Using previous observations taken 9 years earlier
we calculate a proper motion map and combine this with Spitzer 24\,$\mu$m observations of the embedded young stars.\vspace{2mm}
}			
{The combined H$_2$ mosaic shows many new shocks and faint structures in the HH flows. From the proper motion map we find that
most HH objects belong to two major bi-polar HH flows, the large-scale roughly North-South oriented flow from central L1641N
and a previously unseen HH flow in eastern L1641N. Combining the tangential velocity map with the mid-IR Spitzer images, two very likely
outflow sources are found. The outflow source of the eastern flow, L1641N-172, is found to be the currently brightest mid-IR source
in \L1641N and seem to have brightened considerably during the past 20 years. We make the first detection of this source in the
near-IR ($K_S$) and also find a near-IR reflection nebula pointing at the source, probably the illuminated walls of a cone-shaped cavity
cleared out by the eastern lobe of the outflow. Extending a line from the eastern outflow source along the proper motion vector we find that
HH 301 and HH 302 (almost 1\,pc away) belong to this new HH flow.
}			
{}


\maketitle

\section{Introduction}

Very young stars are often associated with molecular bi-polar outflows and jets/shocks that emit both
atomic and molecular lines. At early stages, the outflow and in-fall are believed to occur simultaneously as the
stellar core is formed. These outflows are a good tracer of ongoing star formation, but it is often difficult to exactly localise the central
source which is usually heavily obscured. While single-dish millimetre observations can reveal molecular outflows from
deeply embedded stars at very early stages, they suffer from poor spatial resolution and therefore from source
confusion in crowded regions. Optical surveys (most often H$\alpha$ and [SII]\,6707,6717 nb imaging) instead
often suffer from too high extinction and works best for the blue-shifted lobe of Herbig-Haro (HH) flows with
high inclinations. The shock-heated gas often give rise to H$_2$ emission, and by observing e.g. the 2.12\,$\mu$m line, the cloud extinction 
is less of a problem.  However, the faintness of the emission means that deep images are essential to study these flows in detail.

Lynds 1641 (L1641) is a dark molecular cloud, located in the southern part of our nearest giant molecular cloud Orion A (d$\sim$450\,pc).
It  hosts sites of active star formation (Strom et al. \cite{strom89}), and attention was drawn to the northern part of the cloud, simply
called L1641N, in the mid 80s with the discovery of a molecular outflow (Fukui et al. \cite{fukui86}) at the position of the mid-IR bright
source IRAS 05338-0624.  Fukui et al. (\cite{fukui88}) found well-separated lobes to the
North (blue-shifted) and South (red-shifted) of the IRAS source using CO observations of higher resolution. Further observations have
been made through molecular line studies (e.g. Chen et al. \cite{chen96}, Sakamoto et al. \cite{sakamoto} and Stanke et al. \cite{stanke_06}), Optical
and near-IR surveys (e.g. Strom et al. \cite{strom89}, Hodapp \& Deane \cite{hodapp93}, Chen et al. \cite{chen93}) and in the mid-IR (G\aa lfalk et
al. \cite{galfalk06_1}).

L1641N is presently the most active site of low-mass star formation
in the L1641 molecular cloud and has one of the very highest concentrations of HH objects known anywhere in the sky (Reipurth el al. \cite{reipurth}).
HH shocks in these flows have been observed in both the near-IR 2.12 $\mu$m line of H$_2$ (Davis et al. \cite{davis}, Stanke et al. \cite{stanke}) and
in the optical  lines H$\alpha$ and the [SII] doublet (Reipurth et al. \cite{reipurth}, Mader et al. \cite{mader}).
In a forthcoming paper (G\aa lfalk et al. \cite{galfalk06_1}) we study the young stellar population in
L1641N using ISO, Spitzer, ground based photometry and spectroscopy. In this contribution we instead focus on the HH flows using
our recent 2.12\,$\mu$m H$_2$ observations (which are the deepest and highest resolution observations yet of the HH flows in L1641N).
We have mapped the proper motions of shock heated H$_2$ and related these flows to embedded young stars seen in the mid-IR.
Throughout the paper we use the notation {\it shocks} as short for {\it shock heated regions}. 

\section{Observations and reductions}

\subsection{Narrow band 2.12\,$\mu$m and $K_S$ band imaging}

We have made deep observations of the shocks and jets in L1641N using the 2.12\,$\mu$m S(1) line of H$_2$. Since this line is close to the
centre of our $K_S$ filter (2.14\,$\mu$m) we used this broad band filter to sample the continuum. This is of course a much quicker (although
probably not as accurate) method than using equal amounts of exposure time with a continuum nb filter to one side of the line. It also provides a
well defined band for point source photometry of the outflow sources.
The observations were made on two photometric nights at the Nordic Optical Telescope (NOT) on Dec 13-15 2005 with an average seeing of
0.75\,$\arcsec$ (0.60--0.85\,$\arcsec$) using NOTCam with the newly installed science-grade array. NOTCam is an HAWAII 1024x1024x18.5$\mu$m pixels
HgCdTe array with a field-of-view of 4\farcm 0\,$\times$\,4\farcm 0.
Two positions were selected
on the basis of including as many interesting jets and Herbig-Haro objects as possible in order to relate these to embedded
outflow sources. Both regions include central L1641N which thus has an increased total exposure time. The first
deep field is centred on a position ($05^{h}36^{m}24.07^{s}$, $-06^{\circ}23\arcmin01.9\arcsec$, Epoch 2000) close to the brightest
mid-IR source in L1641N (L1641N-172, see G\aa lfalk et al. \cite{galfalk06_1}). The second
deep field, centred on ($05^{h}36^{m}13.94^{s}$, $-06^{\circ}20\arcmin52.5\arcsec$, Epoch 2000) images the NW part of L1641N but also
overlaps the first mosaic as the central part of L1641N is included in both fields.

All NOTCam observations are differential, meaning that the bias frame and dark current are removed automatically through on-off calculations. A median
sky is subtracted from the target (with equal exposure time) and in these observations, where we do not have any really extended
objects, we use small-step dithering between each exposure and calculate a median sky from the on frames themselves. The flat-fielding is also
differential (sky-flats observed with some time delay and subtracted). Besides the usual reduction steps of near-IR imaging, we have used our
NOTCam model (G\aa lfalk \cite{notcamdist}) to correct for image distortion and some other in-house routines (written in IDL) to find and remove
bad pixels, shift-add images and to remove all the dark stripes that results from lowered sensitivity after a bright source has been read out of
the detector array - this could go unnoticed for normal imaging but when a deep field is made with a lot of overlaps  a complicated pattern may result
(especially after distortion correction) that has to be corrected for in order to be able to keep a high contrast in the mosaics.

The individual exposure times are 14.4\,s and 60\,s for the $K_S$ band and 2.12\,$\mu$m nb filter respectively. In the final mosaics, including
both nights, total exposure times are 2880\,s and 10500\,s for $K_S$ and 2.12\,$\mu$m respectively at position 1 and
1440\,s and 6000\,s respectively at position 2. In the central region of L1641N, where both images overlap we thus get total exposure
times of 4320\,s  and 16500\,s respectively in the two filters.
We reach a 3$\sigma$ H$_2$ surface brightness limit of 4.8\,$\times$\,10$^{-17}$\,erg\,s$^{-1}$\,cm$^{-2}$\,arcsec$^{-2}$ and detect point-sources
down to K$_S$\,$\sim$\,19.5\,mag.

\subsection{Additional ground based observations}

\subsubsection{First epoch 2.12\,$\mu$m H$_2$ imaging}

In order to map the proper motions of the jets and HH objects we needed a previous epoch of 2.12\,$\mu$m observations, with
a large enough time span to show the proper motions with reasonable accuracy, but at the same time with high enough quality to show as many
of our recently observed H$_2$ objects as possible given that the new epoch is a deep field. We would like to thank Stanke et al. for
providing us with the much needed first epoch observations, that were carried out on Dec 26 1996 using the Omega-Prime camera
on the Calar Alto 3.5\,m telescope. For details of these observations we refer to Stanke et al. (\cite{stanke}). Using this mosaic we thus
get a time span of almost exactly 9 years, and since the same type of detector array was used for both epochs the quality is comparable, except
for our much longer total exposure time (obtainable by only imaging L1641N).

\subsubsection{Optical imaging ($I$ band)}

Optical observations are very useful in order to clearly show reflection nebulosity on our side of the cloud.
We have made $I$ band observations on 04 December 2003 using the ALFOSC (Andalucia Faint  Object Spectrograph and Camera) on the Nordic
Optical Telescope (NOT). This instrument has a $2048 \times 2048$ CCD and at a PFOV of 0\farcs 188/pixel it has a FOV of
about 6\farcm 4\,$\times$\,6\farcm 4. A $5\times5$ mosaic was made with individual exposure times of 60\,s using step sizes
of $23\arcsec$ and $30\arcsec$ in RA and Dec respectively, leading to a total exposure time of 25 minutes throughout most of the mosaic.

\subsection{Spitzer Space Telescope}

Spitzer carries a 85\,centimetre cryogenic telescope and three  science instruments, one
of these is the Multi-band Imaging Photometer for Spitzer (MIPS) that contains three separate detector arrays, making simultaneous observations
possible at 24, 70, and 160 $\mu$m. Another instrument is the Infrared Array Camera (IRAC), providing simultaneous 5\farcm 2\,$\times$\,5\farcm 2 images
in four channels, centred at 3.6, 4.5, 5.8 and 8.0 $\mu$m. Each channel is equipped with a 256 $\times$ 256 pixel detector array with a pixel size
of about 1\farcs 2\,$\times$\,1\farcs 2.

The Spitzer data used in this paper was obtained from the Spitzer Science Archive using the Leopard software, and all data had been reduced to
the Post-Basic Calibrated Data (pbcd) level. We used MIPS data from the program
``A MIPS Survey of the Orion L1641 AND L1630 Molecular Cloud'' (Prog.ID 47) and IRAC data from the program
``An IRAC Survey of the L1630 and L1641 (Orion) Molecular Clouds'' (Prog.ID 43), both with G.\,Fazio as the P.I.
We used the 24\,$\mu$m observations of the MIPS program. These observations cover a much
larger region than we need, however, L1641N is covered in the giant mosaic. The 24\,$\mu$m camera has a resolution of 128$\times$128 pixels and
a pixel size of 2.55$\arcsec$. In order to plot Spitzer contours on our ground-based observations, although very different resolutions, we
matched all stars seen in the Spitzer 24\,$\mu$m mosaic with our $K_S$ band observations and corrected for differences in FOV, distortion and
image rotation.

Even though the IRAC data had been reduced to the pbcd level there was still a lot of artefacts, background variations and varying orientation
in the mosaics. For this reason we used our own in-house routines to reduce the data further (this was possible since we had eight overlapping
mosaics to merge in each channel). These additional reductions include cosmic ray removal, de-striping (most likely pick-up noise), removing the
large background variations between regions in the same mosaic and between mosaics, marking bad pixels not to be used further, removing
ghost-effects from bright sources, de-rotation, shifting, adding and making a composite colour image. This required some work, but since we only
needed to reduce the L1641N region (and these mosaics are huge) we could concentrate on just that part - which because of all mosaics having
different rotations meant first finding L1641N in each one and cropping them to manageable sizes.

\section{Proper motion calculations and accuracy}

In order to warp the first epoch images to match the second epoch we start by measuring the positions of all stars visible in both
observations by fitting a two-dimensional elliptical Gaussian equation to each star, including rotation. Writing the equation as

\begin{equation}
	\label{eq_gauss1}
	F(x,y) = A_{0} + A_{1} \cdot e^{-U/2}
\end{equation}

and using the elliptical function

\begin{equation}
	\label{eq_gauss2}
	U = (x^{\prime}/a)^{2} + (y^{\prime}/b)^{2}
\end{equation}

we can include clockwise rotation $\alpha$ with the centre at ($x_{c}$, $y_{c}$) and write $x^{\prime}$ and $y^{\prime}$ as

\begin{equation}
	\label{eq_gauss3}
	x^{\prime} = (x-x_{c}) \cdot cos \, \alpha - (y-y_{c}) \cdot sin \, \alpha
\end{equation}

\begin{equation}
	\label{eq_gauss4}
	y^{\prime} = (x-x_{c}) \cdot sin \, \alpha + (y-y_{c}) \cdot cos \, \alpha
\end{equation}

A gradient-expansion algorithm is used to compute a non-linear least squares fit to Eq.\ref{eq_gauss1} for the parameters
$\alpha$, $a$, $b$,  $A_{0}$, $A_{1}$ and more importantly the centre of the PSF (x$_{c}$, y$_{c}$). This method has proven
to give very accurate sub-pixel positions of stars, accounting for both seeing variations and elliptically shaped stars
due to the optics in the telescope system (image distortion).	

Assuming that the stars, on average, have negligible proper motions and denoting image coordinates in the first epoch (x$_{0}$, y$_{0}$) and
second epoch (x$_{1}$, y$_{1}$) we then use a least squares estimate to fit the following polynomial transformation functions:

\begin{equation}
	\label{eq_xfit}
	x_{1} = \sum_{i=0}^{2} \sum_{j=0}^{2} A_{i,j} \cdot x_{0}^{j} \cdot y_{0}^{i}
\end{equation}

\begin{equation}
	\label{eq_yfit}
	y_{1} = \sum_{i=0}^{2} \sum_{j=0}^{2} B_{i,j} \cdot x_{0}^{j} \cdot y_{0}^{i}
\end{equation}

Even though differences in distortion, field of view and possible slight image rotation between the two epochs have been corrected for using
this method there are several sources of uncertainty related to the geometry and physics of both the HH objects and the reference stars themselves.
Some fast moving stars can have proper motions comparable to that of slow moving shocks and thus increase the uncertainty of the image matching
in parts of the image with few stars. The shocks themselves change shape with time, some much more than
others and for some objects the proper motion varies a lot within the shock itself. Another complicated situation is encountered for HH objects that
are extended along the direction of motion that also lack bright knots that could have been used for easy measurements.

Using the two matched epoch images each proper motion is calculated by sub-pixel shifting the second epoch image so that a shock (or part of a shock
in cases where the proper motion varies within the shock) overlap in both images. This is done manually in steps of 0.1 pixels along both axes while
blinking both images at different frame rates as well as simultaneously displaying them in different colour channels in another
window (first epoch in red colours and second epoch in green). When a shock overlaps itself it appears to be stationary and no colour separations
are seen, the positional shifts are estimated to an accuracy of $\pm$\,0.2 pixels in both directions. Another way to do this would be by using
cross-correlation, but this would however not work for shocks close to stars. Another advantage with the manual method is that we can measure
proper motions for shocks that have varying velocities along the shock or choose a point-like or sharp features that is especially easy to follow.

\begin{figure*}
	\centering
	\includegraphics[width=18cm]{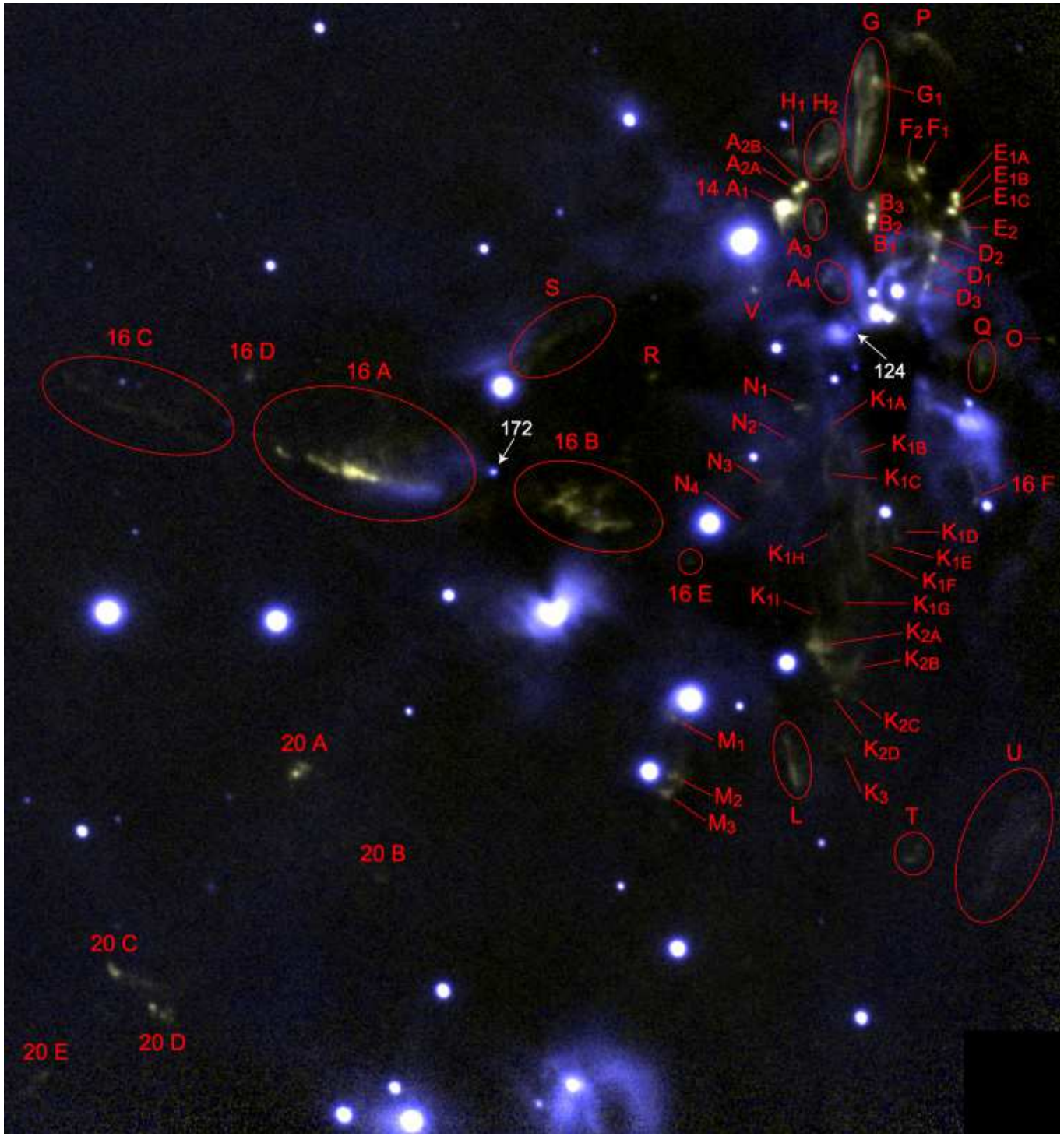}
	\caption{Deep K$_S$ (blue) and 2.12\,$\mu$m H$_2$ (yellow) colour composite of the central and South-East region of
	L1641N including names of all visible H$_2$ features. Most of the features in this Figure belong to group 14, for clarity
	the prefix for these objects have been dropped. Features belonging to the two other groups (16 and 20) are clearly marked with their group name.
	The field shown has a size of 3\farcm 64\,$\times$\,3\farcm 90 and
	is centred on $05^{h}36^{m}24.07^{s}$, $-06^{\circ}23\arcmin01.9\arcsec$ (epoch 2000) - close to the eastern outflow source.
	The two arrows indicate the suggested outflow sources (L1641N-124 and 172).
	}
	\label{Shocks1}
 \end{figure*}

\begin{figure*}
	\centering
	\includegraphics[width=18cm]{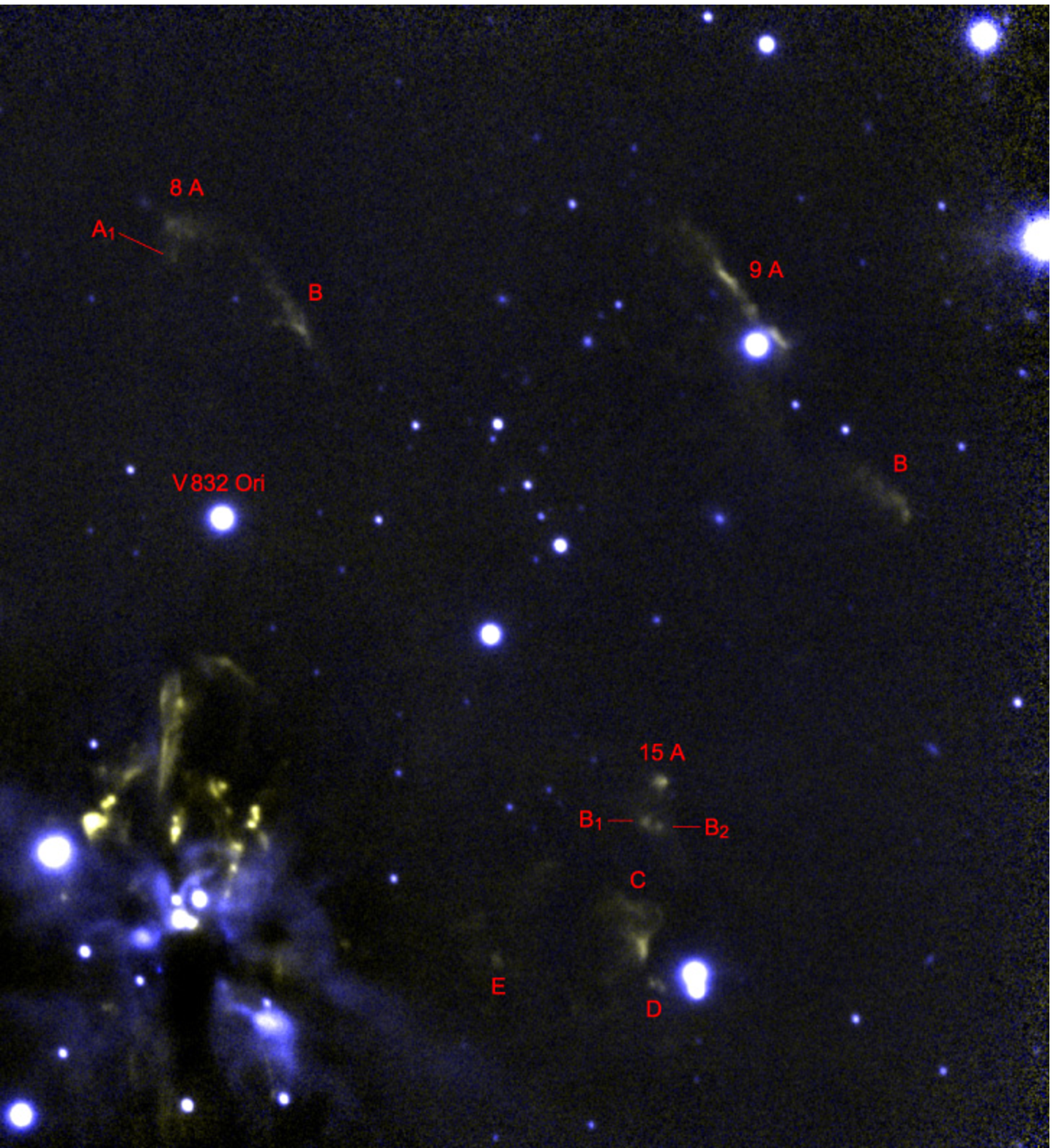}
	\caption{Deep K$_S$ (blue) and 2.12\,$\mu$m H$_2$ (yellow) colour composite of the central and North-West region of
	L1641N including names of all visible H$_2$ features.
	The field shown has a size of 3\farcm 84\,$\times$\,4\farcm 18 and
	is centred on $05^{h}36^{m}13.94^{s}$, $-06^{\circ}20\arcmin52.5\arcsec$ (Epoch 2000).
	}
	\label{Shocks2}
 \end{figure*}

\section{Results and discussion}

Our two 2.12\,$\mu$m H$_2$ deep fields are presented as colour composite images in Figures\,\ref{Shocks1} and \ref{Shocks2} with the 2.12\,$\mu$m
observations as yellow and the continuum ($K_S$ band) as blue. Using an appropriate intensity scaling between the line and continuum images and
by adding a constant to one of the images to make the sky background equal is an efficient way to separate H$_2$ emission from continuum
features without the need to look back and forth at the line and continuum images.
We have continued the H$_2$ shock naming scheme started by Stanke et al. (\cite{stanke}) and added newly discovered objects using
letters (e.g. 14\,R--V) and objects resolved into several components or newly discovered components by sub-group numbers and letters
(e.g. 14\,F is separated into 14\,F$_1$ and F$_2$, 14\,E$_1$ is seen as the three components E$_{1A}$--E$_{1C}$).
In Fig.\ref{Shocks1} most shocks belong to group\,14 and only the sub-groups (A-V) are shown for clarity, but for the H$_2$ features in groups
16 and 20 the group number is shown for each object in order to avoid confusion. Group\,8, seen in Fig.\ref{Shocks2} is part of the same flow as
the group\,14 shocks. In the following, when we refer to the embedded stars themselves by a number (e.g. L1641N-172, or No.\,172 for short) we use the
source numbering found in our forthcoming paper on the IMF in L1641N (G\aa lfalk et al. \cite{galfalk06_1}). Our full catalogue contains a
total of 216 sources, whereas Table~\ref{tab_stars} only lists the small selection of these sources that are discussed in this paper.

\begin{table*}
	\caption{Proper motion results. Column\,1, the shock designations, are given in
	Figures\,\ref{Shocks1} and \ref{Shocks2}. Columns 2--4 give the angular distance
	each shock has moved between the two epochs (almost 9 years apart). Columns 5-7 are the proper
	motions and finally Columns 8--9 the (tangential) velocities and positional
	angles. Uncertainties are given at the bottom of the table (except for the PA
	where this is given in each row).
	}
	\label{propermot}
	\begin{tabular}{lrrrrrrrll}
	  \hline
        \noalign{\vspace{0.5mm}}
        Shock & $\Delta_{\alpha}$ & $\Delta_{\delta}$ & $\Delta_{tot}$ & $\mu_{\alpha}$ & $\mu_{\delta}$ & $\mu_{tot}$ & vel & PA$^{\mathrm a}$ & Comment \\
             & (\arcsec) & (\arcsec) & (\arcsec) & (mas/yr) & (mas/yr) & (mas/yr) & (km/s) & (deg) & \\
	\noalign{\vspace{0.5mm}}
 	(1)  & (2)	 & (3)	     & (4)	 & (5)	    & (6)      & (7)	  & (8)	   & (9)   & (10) \\
 	\noalign{\vspace{0.5mm}}
        \hline

	\noalign{\vspace{1.0mm}}

	~~8\,A &	-0.19 &	+0.61 &	0.64 & -21 & 68 & 71 &	152 & 343 $\pm$  6 & \\
	~~~~\,B &	-0.07 &	+0.31 &	0.31 & -8 & +34 & 35 &	75  & 347 $\pm$ 12 & \\
	\noalign{\vspace{1.0mm}}
	14\,G &	+0.14 &	+0.94 &	0.95 & +16 & +105 & 	106 &	227 & ~~~~9 $\pm$  4 & North (fastest) part \\
	~~~~\,G$_{1}$ &	+0.00 &	+0.19 &	0.19 & 0 & +21 & 21 &	45  & ~~~~0 $\pm$ 19 & \\
	~~~~\,H$_{1}$ &	+0.07 &	+0.24 &	0.25 & +8 & +26 & 27 &	58  & ~~17 $\pm$ 14 & \\
	~~~~\,K$_{1A}$ &	+0.00 &	-0.14 &	0.14 &  0 & -16 & 16 &	34  & 180 $\pm$ 27 & \\
	~~~~\,K$_{1C}$ &	+0.00 &	-0.16 &	0.16 &  0 & -18 & 18 &	39  & 180 $\pm$ 22 & \\
	~~~~\,K$_{1F}$ &	+0.24 &	-0.42 &	0.48 & +26 & -47 & 54 &	115 & 151 $\pm$  8 & \\
	~~~~\,K$_{2A}$ &	+0.12 &	-0.47 &	0.49 & +13 & -53 & 54 &	116 & 166 $\pm$  8 & \\
	~~~~\,K$_{2B}$ &	+0.09 &	-0.89 &	0.90 & +11 & -100 & 100 & 214 & 174 $\pm$  4 & \\
	~~~~\,K$_{2D}$ &	+0.12 &	-0.89 &	0.90 & +13 & -100 & 101 & 215 & 173 $\pm$  4 & \\
	~~~~\,L &	+0.05 &	-0.52 &	0.52 & +5  & -58 & 58 &	124 & 175 $\pm$  7 & \\
	~~~~\,M$_{2}$ &	+0.14 &	-0.26 &	0.30 & +16 & -29 & 33 &	70  & 151 $\pm$ 13 &  \\
	~~~~\,M$_{3}$ &	+0.14 &	-0.26 &	0.30 & +16 & -29 & 33 &	70  & 151 $\pm$ 13 & \\ 
	~~~~\,P &	+0.07 &	+0.68 &	0.69 & +8 & +76 & 77 &	163 & ~~~~6 $\pm$  5 & \\
	~~~~\,R &	+0.59 &	-0.26 &	0.64 & +66 & -29 & 72 & 153 & 114 $\pm$  6 & \\
	~~~~\,V &	-0.47 &	-0.42 &	0.63 & -53 & -47 & 71 &	151 & 228 $\pm$  6 & \\
	\noalign{\vspace{1.0mm}}
	16\,A &	+0.31 &	+0.07 &	0.31 & +34 & +8 & 35 &	75  & ~~77 $\pm$ 11 & \\
	~~~~\,B &	-0.40 &	+0.00 &	0.40 & -45 & 0 & 45 &	95  & 270 $\pm$  8 & \\
	~~~~\,D &	+0.35 &	-0.28 &	0.45 & +39 & -32 & 50 &	108 & 129 $\pm$  9 & \\

	\noalign{\vspace{1.0mm}}
        \hline
        \noalign{\vspace{0.5mm}}
	   &  $\pm$ 0.05 & $\pm$ 0.05 & $\pm$ 0.07 & $\pm$ 6 & $\pm$ 6 & $\pm$ 8 & $\pm$ 16 & & \\
	\noalign{\vspace{0.5mm}}
	\hline

	\end{tabular}

	\begin{list}{}{}
		\item[$^{\mathrm{a}}$] Position Angle: North 0\,$\degr$, East 90\,$\degr$, South =180\,$\degr$, West 270\,$\degr$.
	\end{list}
\end{table*}

\begin{table}
\caption{Selected L1641N sources$^{\mathrm a}$.}
\label{tab_stars}
\begin{tabular}{lcccl}
\hline
\noalign{\vspace{0.5mm}}
No. & RA & Dec & K$_S$ & Comment \\
      & (2000) & (2000) & (mag) & \\

\noalign{\vspace{0.5mm}}
\hline

\noalign{\vspace{1.0mm}}

61  & 05:36:10.44 & -06:20:01.5 & 10.85 $\pm$ 0.01 & \\
110 & 05:36:17.86 & -06:22:28.6 & 15.28 $\pm$ 0.08 & \\
114 & 05:36:18.48 & -06:20:38.7 & 11.00 $\pm$ 0.01 & V832 Ori \\
115 & 05:36:18.49 & -06:22:13.6 & ... & \\
116 & 05:36:18.81 & -06:22:10.6 & ... & \\
{\bf 124} & 05:36:19.50 & -06:22:12.1 & 16.40 $\pm$ 0.15 & Central flow \\
145 & 05:36:21.87 & -06:23:30.1 & 10.71 $\pm$ 0.02 & \\
{\bf 172} & 05:36:24.61 & -06:22:41.7 & 15.67 $\pm$ 0.04 & Eastern flow \\

\noalign{\vspace{1.0mm}}
\hline

\end{tabular}
	\begin{list}{}{}
	\item[$^{\mathrm{a}}$] All data and source numbers in this Table are from our forthcoming paper on the IMF in L1641N.
	The full catalogue contains 216 sources. 
	\end{list}
\end{table}

The results of our proper motion measurements throughout L1641N are presented in Table~\ref{propermot}.
As shown in the Table the tangential velocities (assuming a distance of 450\,pc) are in the range 30--230\,km\,s$^{-1}$.
The lower value also roughly represent the detection limit.
Since the radial velocities of these H$_2$ features are unknown the measurements are lower limits of the real three
dimensional velocities, yet some of the proper motions suggest velocities in excess of 200\,km\,s$^{-1}$, much higher than
the dissociation speed limit of H$_2$ in molecular shocks of $\sim$25 and $\sim$50\,km\,s$^{-1}$ for J- and C-type shocks, respectively (Smith \cite{smith94}).
This may at first seem like a contradiction, given that all the proper motions were measured and thus clearly seen in H$_2$ emission
(suggesting low shock velocities). Proper motions are however not necessarily a direct measurement of the shock speed for several reasons.
Proper motions measure the pattern speed of these features, but for example in the case of flow variability (Raga \cite{raga}) the pre-shock gas could
also be non-stationary resulting in internal working surfaces in the flow, bounded by a leading and a trailing reverse shock, possibly with
only marginally faster material moving into somewhat slower gas and therefore much lower shock velocities than suggested directly from
the pattern speed.
Also, since the shock speed is locally determined by the velocity perpendicular to the shock surface, in a bow shock there is a range of
shock velocities involved from the fast shock at the apex to the much slower shocks in the wings. This then means that even if a bow shock
is much faster than the dissociation speed at its apex, there could still be bright H$_2$ emission coming from the wings
(see e.g. the bow shock models in Smith et al. \cite{smith03}). Several other 2.12\,$\mu$m H$_2$ proper motion surveys of star-formation regions
have also resulted in tangential velocities much higher than the H$_2$ dissociation speed limit for both J- and C-shocks (Coppin
et al. \cite{coppin}; Micono et al. \cite{micono}; Lee et al. 2000).

\begin{figure*}
	\centering
	\includegraphics[width=18cm]{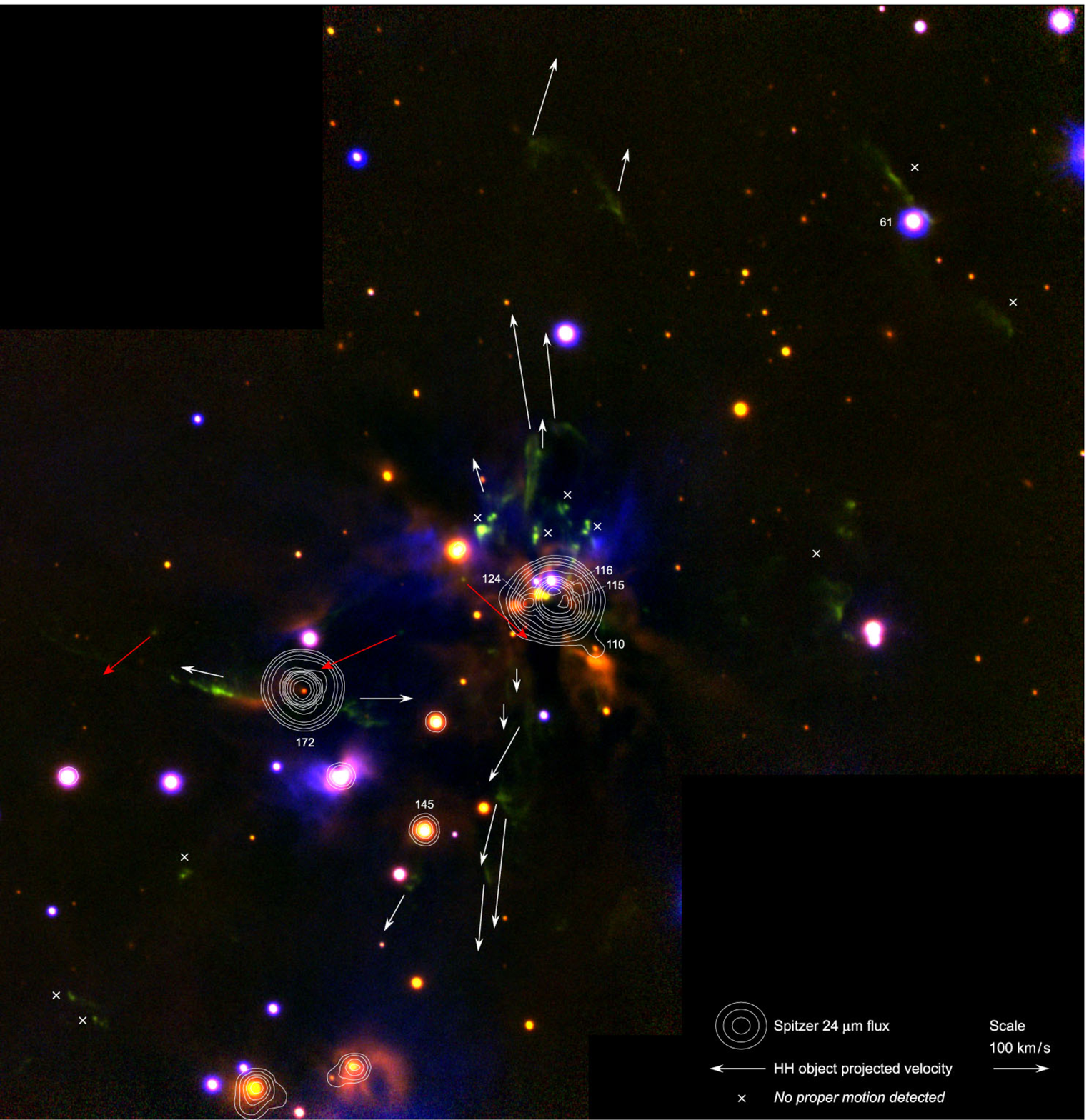}
	\caption{Flow chart of Herbig-Haro objects. Proper motions are indicated by arrows overplotted on a colour composite of
	our $K_S$ (red), 2.12\,$\mu$m H$_2$ (green) and $I$ (blue) mosaics. White arrows indicate HH objects that are suggested by
	their proper motions to be part of either the major North-South oriented HH flow emanating from central L1641N or
	part of the proposed new HH flow. Red arrows are used for single HH objects with uncertain origin. Bright HH objects for
	which we detect no movement are marked with an x. Spitzer 24\,$\mu$m contours are also overplotted, revealing the most
	embedded and young stars, responsible for the jets and shocks. Numbers refer to deeply embedded sources as given in
	G\aa lfalk et al. (\cite{galfalk06_1}) and Table~\ref{tab_stars}. The field of view of this mosaic is roughly
	6\farcm 1\,$\times$\,6\farcm 3.
	}
	\label{Flowplot}
 \end{figure*}

\begin{figure*}
	\centering
	\includegraphics[width=18cm] {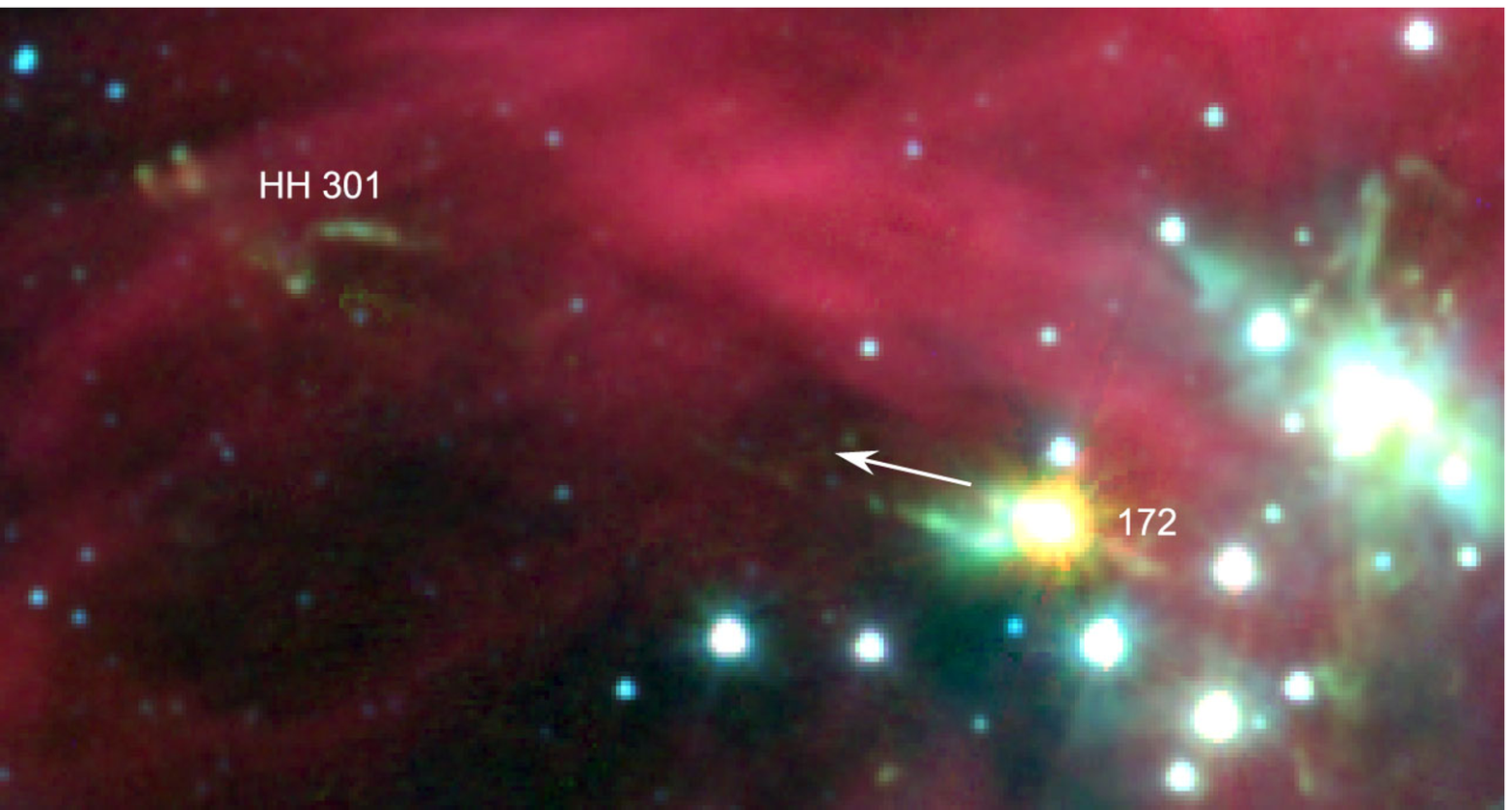}
	\caption{Spitzer colour mosaic of the eastern flow using IRAC channels 3.6\,$\mu$m (blue), 4.5\,$\mu$m (green) and 8.0\,$\mu$m (red).
	L1641N-172 is very bright (even saturated) in the 4.5 and 8.0\,$\mu$m images. Shocked H$_2$ is clearly seen in the 4.5\,$\mu$m
	band close to L1641N-172 at the same position as in our 2.12\,$\mu$m observations. Following the direction of proper motion
	(shown by the arrow) leads to a group of H$_2$ shocks (HH 301) outside of our 2.12\,$\mu$m mosaic (see Fig.\ref{Shocks1}).
	The field of view is 6\farcm 1\,$\times$\,3\farcm 3. North is up and East is left in all Figures.
	}
	\label{Eastern}
 \end{figure*}

In Fig.\ref{Flowplot} as an overview we present all proper motions drawn on a colour composite mosaic ($K_S$, 2.12\,$\mu$m H$_2$ and $I$) of
the full surveyed region. Spitzer 24\,$\mu$m flux contours are also overplotted since cold, deeply embedded young stars are bright at these
wavelengths. The projected velocity vectors are drawn using two colours, white arrows relate to H$_2$ features of the major flows
while red arrows are used for features with unknown origins. The general impression from all shocks with measurable proper motions, in
combination with the Spitzer 24\,$\mu$m contours, is that most shocks are either part of the bi-polar large scale North-South moving flow
emanating from central L1641N or a previously undetected roughly East-West bi-polar HH flow originating at the other
strong 24\,$\mu$m source, also clearly seen in our $K$ band image.
In addition to this, there are a large amount of H$_2$ features where no motion is detected at our present
resolution (pixel size of 0.40\,$\arcsec$ and 0.23\,$\arcsec$ for the epochs) and time span (9 years)
meaning that they are most likely shocks with small projected velocities (slower
than about 30\,km\,s$^{-1}$). These features have been marked with an x in Fig.\ref{Flowplot}.
While the morphology of features like 14~D$_1$--D$_3$ and E$_1$ (knots lying almost exactly along a
line through the very central of L1641N) suggest a HH flow nature, other features (e.g. groups 9, 15 and 20) lack this morphology and could
possibly be PDR-style fluorescently-excited layers and filaments in the surrounding cloud.

\subsection{The eastern flow}
Spitzer observations of the eastern bi-polar flow using IRAC (3.6, 4.5 and 8.0\,$\mu$m) are shown in Fig.\ref{Eastern}.
The field of view extends beyond
the eastern border of our 2.12\,$\mu$m mosaic. Following the proper motion of shocked H$_2$ in jet 16\,A close to the mid-IR
bright source L1641N-172 (as shown by the arrow) leads to a group of objects that also have morphologies consistent with bow shocks. This is
in fact HH 301 and as can be seen in the wide-field H$_2$ mosaic of Stanke et al. (\cite{stanke}), continuing along
this line leads to HH 302.
It has been seen in several other Spitzer surveys (e.g. Noriega-Crespo et al. \cite{noriega04} and Harvey et al. \cite{harvey06}) that IRAC
channel 2 (4.5\,$\mu$m) is very efficient in detecting Herbig-Haro objects. The reason for this is partly that the spectral response function is
highest in this channel, but there are at least two more contributing factors. Between approximately 4--5\,$\mu$m there are many vibrational
and rotational H$_2$ emission lines, these have been modeled by Smith \& Rosen (\cite{smith}) for all IRAC bands using three-dimensional hydrodynamic
simulations of molecular jets. The strongest integrated H$_2$ emission is predicted to arise from band 2 because of rotational transitions.
For the typical conditions of low-mass outflows, pure-rotational transitions like S(11)--S(4) (4.18--8.02\,$\mu$m) can actually be much brighter
than the standard 2.12\,$\mu$m H$_2$ line (Kaufman \& Neufeld \cite{kaufman}). Channel 2 is also the most ``PAH-free'' band of IRAC, greatly enhancing
its usefulness as a HH tracer, as opposed to the 5.8 and 8.0\,$\mu$m channels (see Fig.\ref{Eastern}) which include bright, diffuse
emission from polycyclic aromatic hydrocarbons (PAHs), hiding the shock-excited H$_2$ features of the HH flows.

Both HH 301 and HH 302 are therefore very likely part of the eastern HH-flow, originating from L1641N-172, which
we confirm as the outflow source using
proper motions, near-IR 2.12\,$\mu$m H$_2$ morphology and mid-IR photometry at 3.6, 4.5, 5.8, 8.0 and 24\,$\mu$m.
HH 302 is located $\sim$\,6.3$\arcmin$ away from L1641N-172 (corresponding to 0.83\,pc at the assumed distance of 450\,pc)
suggesting that this is
a large-scale flow. In the opposite direction in the same wide-field mosaic there is another HH object (called 19 in Stanke et al. \cite{stanke}) that
could also be part of the flow, although proper motions are required to confirm this.

Note that the proper motion vector of the H$_2$ feature in the counter-flow 16\,B (West of the outflow source) only refers to its fast moving
northern part. There seems to be very different velocities involved across feature 16\,B and
the proper motion vector may just represent the fastest moving part in a probable cone-like outflow with a similar opening angle as
on the other side of the outflow source.

As can be seen in Fig.\ref{Flowplot} the position of the outflow source, L1641N-172, coincides very well in the $K_S$ and 24\,$\mu$m
observations. Looking at the two epochs we use for proper motion calculations this source has brightened considerably in the last
9 years (and comparing IRAS and ISO observations it seems to have brightened a lot over the decade before that as well). Since this
source is not detected in the first epoch K$_S$ mosaic (limiting magnitude K$_S$\,$\sim$\,17) we can however only give a lower limit of
1.5 mag. for the increased brightness between the two epochs.
 L1641N-172 is extremely bright in the mid-IR and saturates Spitzer at 4.5, 5.8 and 8.0\,$\mu$m.
In our new $K_S$ images there is also clearly a cone-shaped reflection nebula pointing at the outflow source, which is apparently
lit up by the now much brighter source. The shape of the nebula suggests that it could be the illuminated walls of a cone-shaped
cavity cleared out by this previously undetected flow.

Both this new bi-polar HH flow and its suggested outflow source (L1641N-172) are supported by recent CO observations
(Stanke et al. \cite{stanke_06}, to be published) which reveal a bi-polar CO outflow centred on this source, with a blue-shifted lobe roughly to the
East (B-E) and a red-shifted lobe to the West (R-E), in agreement with our proper motions.

HH 298 is an East-West chain of three HH knots about 70\,$\arcsec$ long, originally reported in Reipurth et al. \cite{reipurth}.
We note that there seems to be some general confusion about this object in several publications. Reipurth et al. states knot A as the
brightest knot in H$\alpha$ and [SII], which is the westernmost knot as is also marked in their Fig.5, but gives its coordinates as the easternmost
knot (which is knot C in their text). In Mader et al. (\cite{mader}) HH 298 A is given (in their Fig.9) as a different source that is not
close to either knot HH 298 A, B or C but instead close to our discovered outflow source of the eastern HH flow, L1641N-172.
They also suggest N23 of Chen et al. (\cite{chen93}) as the outflow source for HH 301 and HH 302. However, we do not
detect any point source at that position in our deep field
(it is probably a bright H$_2$ knot in the 16\,A jet that Chen et al. detected in the $K$ band as N23).
Only the westernmost (and optically brightest) knot of HH 298 has an IR counterpart. It corresponds to
shocks E$_{1A}$--E$_{1C}$ which are the three outermost knots in a well-defined chain of knots towards the NNW from
central L1641N. Thus, the three knots of HH 298 (lying on a E-W line) cannot be part of the same flow, as seen from geometry.

\subsection{The central North-South flow}

Comparing the velocities in this flow there are many knots with no measurable proper motions near the central source, while
further out features with large proper motions are seen. This is roughly consistent with a hubble flow, though obviously
orientation will play a major role.
Using two bright, well-defined bow shocks
(14\,K$_{2B}$ and 14\,P) with accurate proper motions, we find that very similar ages are implied (730 and 780 years respectively). This suggests
that they were probably ejected in the same event. Looking at the bow shock
8\,A further to the north (with an implied age of 2200\,years) this seems to be the leading shock front of an earlier event.
Shocks from older outbursts than that have probably moved out of our observed region. Following the mean velocity vector South in a wide field
mosaic (Fig.1 of Stanke et al.\cite{stanke}) we find groups of HH objects (23\,A, B, D, E, F, G) far away from the source. Similarly, following
the velocity vector of the northern flow we find two groups (4\,A and B) of shocks at about the same distance from the outflow source
as 23\,F but on the other side. This is thus a very large-scale outflow extending several parsecs as seen in the IR. There is a bright
star, V832 Ori (L1641N-114), in the north lobe of the central flow (see Fig.\ref{Shocks2}). In our survey of young stars in
L1641N (G\aa lfalk et al. \cite{galfalk06_1}) optical spectra show this to be a young star with T$_{eff}$ = 3800\,K and an IR
excess at 8.0\,$\mu$m
(as seen in our Spitzer photometry). This star does however not show any evidence in our 2.12\,$\mu$m images of being physically
related to the HH flow in any way.

We make the first detection of a previously proposed central outflow source, L1641N-124, in the near-IR (K$_S$ = 16.40 $\pm$ 0.15).
This source is very bright in the mid IR, it saturates Spitzer at 8.0\,$\mu$m and has a high flux at 24\,$\mu$m. The proper motions in the
flow suggest this as the most likely outflow source to the giant, North-South flow. There is a bright reflection nebula next (to the East) to this source.
Chen et al. (\cite{chen93}) have observed this source (called N15 in their paper) in the M band, however, their $H$ and $K$ source is in fact
the nearby reflection nebula since it looks extended and is displaced to the East of the outflow source.

The central region is very crowded with H$_2$ shocks, especially to the north, complicating the situation of a single outflow source.
While L1641N-124 is the most likely major outflow source (most easily seen from the proper motions in the southern part) there are
probably also outflows from L1641N-115, 116 and 110. An example of this is the very collimated chain of H$_2$ knots pointing out from
a position close to L1641N-115 and 116 to the NNW (objects 14\,D$_1$--D$_3$ and E$_{1A}$--E$_{1C}$). Since no proper motion is detected
in this chain of shocks it is either slow or has a different inclination (more radial than tangential) than the major flow which in that
case suggests a different outflow source that could very well be No.\,115 or 116. It is probably the blue-shifted jet that we see in the
NNW flow since no counter-jet is seen to the SSE.

We cannot rule out (using proper motions or the 24\,$\mu$m flux) the possibility that L1641N-115 or 116 contribute to
the giant, North-South flow. One reason for this is that even though the northern jet-like feature (14\,G) extending out from central L1641N
clearly point away from L1641N-124, this
could be a shock along the wall of the cavity as the flow interacts with the surrounding molecular cloud. In the optical  lines
H$\alpha$ and [SII] (see Reipurth et al. \cite{reipurth}) this jet (HH\,303) is displaced to the West (pointing at 14\,P). Also, while most proper
motions South of L1641N-124 point away from this source, there are several shocks to the North that point away from L1641N-115/116.
Combining IR and optical observations, Reipurth et al. (\cite{reipurth}) suggests that the southern flow is red-shifted (since it is not
seen in the optical) and the northern flow is blue-shifted and seem to stretch 6.3\,pc away including HH\,306-310 in the optical. At the far
South (6.5\,pc away) HH\,61/62 are seen at the southern edge of the L1641N cloud (where the extinction is low) and could form a counter-lobe
of about equal length. This scenario agrees well with the red and blue-shifted lobes seen in CO observations (Fukui et al. \cite{fukui88}).
The flow is probably oriented close to the plane of the sky since its fastest shocks have projected velocities above 200\,km\,s$^{-1}$ as shown in our
proper motion measurements.

\subsection{Other shocks}
In Fig.\ref{Flowplot} we have used red arrows to mark proper motions of isolated H$_2$ knots with unknown
origins (shocks 14\,R, 14\, V and 16\,D). This has been done for completeness of the survey and the names have to do with the fact that they are
located close to the corresponding flows although not part of them. Object 14\,R however has a proper motions that suggests that it could originate
from a source in the central region.
The proper motions and locations of H$_2$ objects 14\,M$_2$ and M$_3$ suggests two possibilities. They could originate from a source in the central
region, but the observations are also in agreement with a bi-polar outflow from L1641N-145 (which is also bright at 24\,$\mu$m) that has
14\,M$_1$--M$_3$ on one side and the chain of shocks 14\,N$_1$--N$_4$ to the other side. Even though the shocks 9\,A and B (HH\,299) are located
very close to the young star L1641N-61 (see Figures \,\ref{Flowplot} and \ref{Shocks2}) this is most likely just a coincidence as the geometry suggests that
they are not physically related. As no proper motions are seen for these two shocks at our present time span and resolution this can however not
be confirmed.

\section{Summary and conclusions}

We have made detailed observations of the Herbig-Haro (HH) flows in L1641N by observing two overlapping deep fields in the
near-IR 2.12\,$\mu$m shock line of H$_2$. Using previous observations, taken 9 years before, we also calculate proper motions for
all shocks with projected velocities faster than about 30\,km\,s$^{-1}$.
Two major bi-polar HH flows are seen and most HH objects are shown to be part of either the large-scale roughly North-South
oriented flow from central L1641N or a previously unseen HH flow in eastern L1641N. By combining this velocity survey with new
Spitzer 24\,$\mu$m observations we have also found the outflow source of the eastern flow and found a candidate outflow
source for the giant central N-S flow that we detect for the first time in the near-IR ($K_S$ band). Extending a line from the eastern
outflow source along the proper motion vector of the eastern lobe we find that HH 301 and HH 302, almost 1\,pc away, most
likely are part of this flow. The outflow source of the eastern flow, L1641N-172, is found to be the currently brightest
mid-IR source in L1641N and seem to have brightened
a lot during the last 20 years. We make the first detection of this source in the near-IR ($K_S$) and also find a near-IR
reflection nebula pointing at the source. This could be the illuminated walls of a cone-shaped cavity cleared out by the eastern lobe of the outflow.
The tangential velocities of the central flow close to the outflow source has a explosion-like structure, with a fast moving bow shock followed by
slower and slower moving shocks closer to the source. We find a bow shock in the
northern and southern lobes that were ejected about 730 and 780 years ago respectively, probably belonging to the same outburst. There is another
strong bow shock further out in the North flow that was ejected in a previous outburst about 2200 years ago.

\begin{acknowledgements}
	The Swedish participation in this research is funded by the Swedish National Space Board.
	This publication made use of the NASA/IPAC Infrared Science Archive, which is operated by the Jet Propulsion
	Laboratory, California Institute of Technology, under contract with the National Aeronautics and Space
	Administration, and data products from the Two Micron All Sky Survey, which is a joint project of the University
	of Massachusetts and the Infrared Processing and Analysis Center/California Institute of Technology, funded
	by the National Aeronautics and Space Administration and the National Science Foundation.
	 We would also like to thank Thomas Stanke for the 2.12\,$\mu$m H$_2$ mosaic we used for the
	first epoch of our proper motion calculations.
\end{acknowledgements}

\end{document}